\documentclass[11pt]{article}
\usepackage{graphicx}  
\usepackage{enumerate}
\usepackage[T1]{fontenc}

\usepackage{caption}
\usepackage{fancyhdr}
\newlength\FHoffset
\fancyheadoffset{\FHoffset}
\usepackage{authblk}

\usepackage{geometry}
\title{\Large\bf Tunable on-chip electro-optic frequency-comb generation at 8 µm wavelength}
\date{}  
\author[1]{Victor Turpaud\footnote{Corresponding Author: victor.turpaud@universite-paris-saclay.fr}}
\author[1]{Thi-Hao-Nhi Nguyen}
\author[1]{Natnicha Koompai}
\author[1]{Jonathan Peltier}
\author[2]{Jacopo Frigerio}
\author[2]{Stefano Calcaterra}
\author[1]{Jean-René Coudevylle}
\author[1]{David Bouville}
\author[1]{Carlos Alonso-Ramos}
\author[1]{Laurent Vivien}
\author[2]{Giovanni Isella}
\author[1]{Delphine Marris-Morini\footnote{Corresponding Author: delphine.morini@universite-paris-saclay.fr}}

\affil[1]{Centre de Nanosciences et de Nanotechnologies, CNRS, Université Paris-Saclay, 91120 Palaiseau, France}
\affil[2]{L-NESS, Dipartimento di Fisica, Politecnico di Milano, Polo di Como, Via Anzani 42, 22100 Como, Italy}

\begin{document}

\maketitle

\begin{abstract}\small 
Dual-comb spectroscopy is a powerful technique to measure optical spectra in a wide spectral range with high-frequency resolution. The development of compact systems operating in the long-wave infrared wavelength range is of high interest for spectroscopic and sensing applications. Amongst the different techniques to obtain optical frequency-combs, electro-optic frequency-comb generation presents major advantages thanks to the tunable repetition rate only limited by the bandwidth of the used electro-optical modulator. However, the development of integrated and efficient electro-optical modulators operating in a wide long-wave infrared spectral band is still at its infancy, and electro-optical frequency-comb has not been demonstrated so far beyond the telecom band. In this work, a Schottky-based modulator embedded in a Ge-rich graded SiGe waveguide is used for electro-optic frequency-comb generation. Considering the limited efficiency of the modulator, harmonically-rich RF signals are used to enhance the generation of comb lines around the optical carrier. Interestingly, this allows us to demonstrate the generation of electro-optical combs spanning over 2.4 GHz around 8 µm wavelength. This paves the way towards fully integrated and tunable mid-infrared electro-optic frequency-comb generation systems.
\end{abstract}


\section{Introduction}
Optical frequency-combs (OFC) are sets of discrete, equally spaced and mutually coherent laser lines. The first observation of such an OFC dates back to 1964 \cite{hargroveLockingHeNe1964}, with the invention of the first mode-locked laser (MLL). Since then, this particular type of spectra aroused great interest in a wide range of applications requiring coherent and/or equally spaced lines in the frequency domain, including precision frequency metrology \cite{udemOpticalFrequencyMetrology2002,leziusSpaceborneFrequencyComb2016}, wavelength division multiplexing (WDM) for telecommunications \cite{huChipbasedOpticalFrequency2021}, radio-frequency (RF) signal processing \cite{torres-companyOpticalFrequencyComb2014} or spectroscopy \cite{denielFrequencytuningDualcombSpectroscopy2020,
picqueFrequencyCombSpectroscopy2019,millotFrequencyagileDualcombSpectroscopy2016}. Regarding this last application, mid-infrared (mid-IR) OFC would allow to address the fundamental absorption frequencies of molecular vibrational and rotational modes \cite{workmanHistoryNearInfraredNIR2007}.\\
Mid-IR OFC has already been demonstrated by mean of quantum cascade lasers (QCLs) \cite{schliesserMidinfraredFrequencyCombs2012,hugiMidinfraredFrequencyComb2012,
taschlerFemtosecondPulsesMidinfrared2021} or electro-optic (EO) modulation in the near-infrared (NIR) followed by nonlinear frequency conversion processes \cite{kowligyMidinfraredFrequencyCombs2020,yanMidinfraredDualcombSpectroscopy2017}. In order to provide mid-IR sensing platforms at a low cost, unlocking the possibilities to enhance food safety, diseases detection or pollutants monitoring, it is of great interest to develop such platforms in integrated photonics \cite{popaIntegratedMidInfraredGas2019}. Amongst the different available integrated mid-IR photonics platforms such as chalcogenide glasses \cite{gutierrez-arroyoOpticalCharacterizationUm2016}, suspended silicon \cite{penadesSuspendedSOIWaveguide2014}, SiGe on Si \cite{grilletLowLossSiGe2013} or Ge on Si \cite{nedeljkovicGermaniumonsiliconWaveguidesOperating2017}, the Ge-rich graded SiGe platform appears to be promising as it enables low propagation losses in a wide spectral range of the mid-IR \cite{marris-moriniGermaniumbasedIntegratedPhotonics2018,montesinos-ballesterGerichGradedSiGe2020}. Finally, recent demonstrations of high-speed mid-IR modulators \cite{montesinos-ballesterMidinfraredIntegratedElectrooptic2022,nguyenGHzElectroopticalSilicongermanium2022} enable the direct generation of electro-optic frequency-combs (EOFC) in the mid-wave infrared (MWIR) and long-wave infrared (LWIR), from 5 to 9 µm wavelength. This approach allows the generation of OFC whose linespacing and shape are as versatile as the electronic waveform-generator used to drive the modulator. In this work, we show that, by mean of a Schottky-based EO modulator, it is possible to generate tunable mid-IR EOFC spanning over 2.4 GHz around a carrier wavelength of 8 µm.

\section{Theory}
\begin{figure}[ht]
	\centering
	\includegraphics[width=\linewidth]{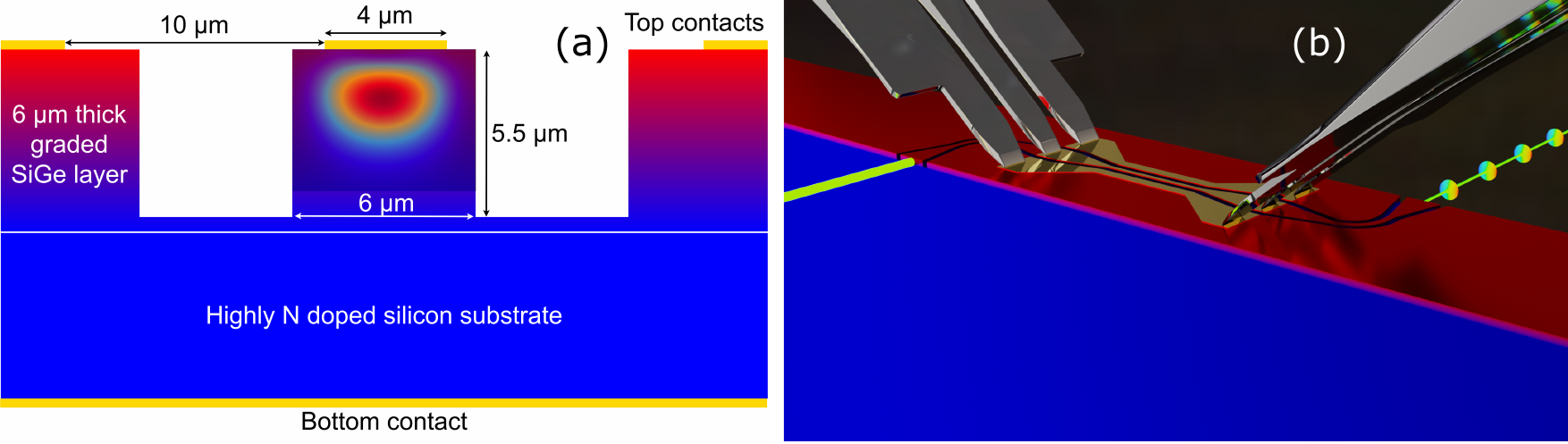}
	\caption{\textbf{EOFC generation on the graded SiGe platform.} \textbf{(a)} Cross-section view of the Schottky-based modulator on graded SiGe used. The germanium concentration is represented in colorscale from 0 (blue, pure silicon) to 1 (red, pure germanium) on both sides. The index gradient, similar to the germanium concentration gradient, allows a light confinement close to the Schottky contact as shown in the centre, ensuring a large overlap between the optical mode and the voltage dependent space charge region. \textbf{(b)} EOFC generation principle. A CW laser source is sent to the modulator, driven by a periodic electrical signal, which modifies both the light amplitude and instantaneous phase and frequency, leading to chirped optical pulses at the output.}
	\label{fig:EOcombPrinciple}
\end{figure}

\begin{figure}[ht]
	\centering
	\includegraphics[width = .8\linewidth]{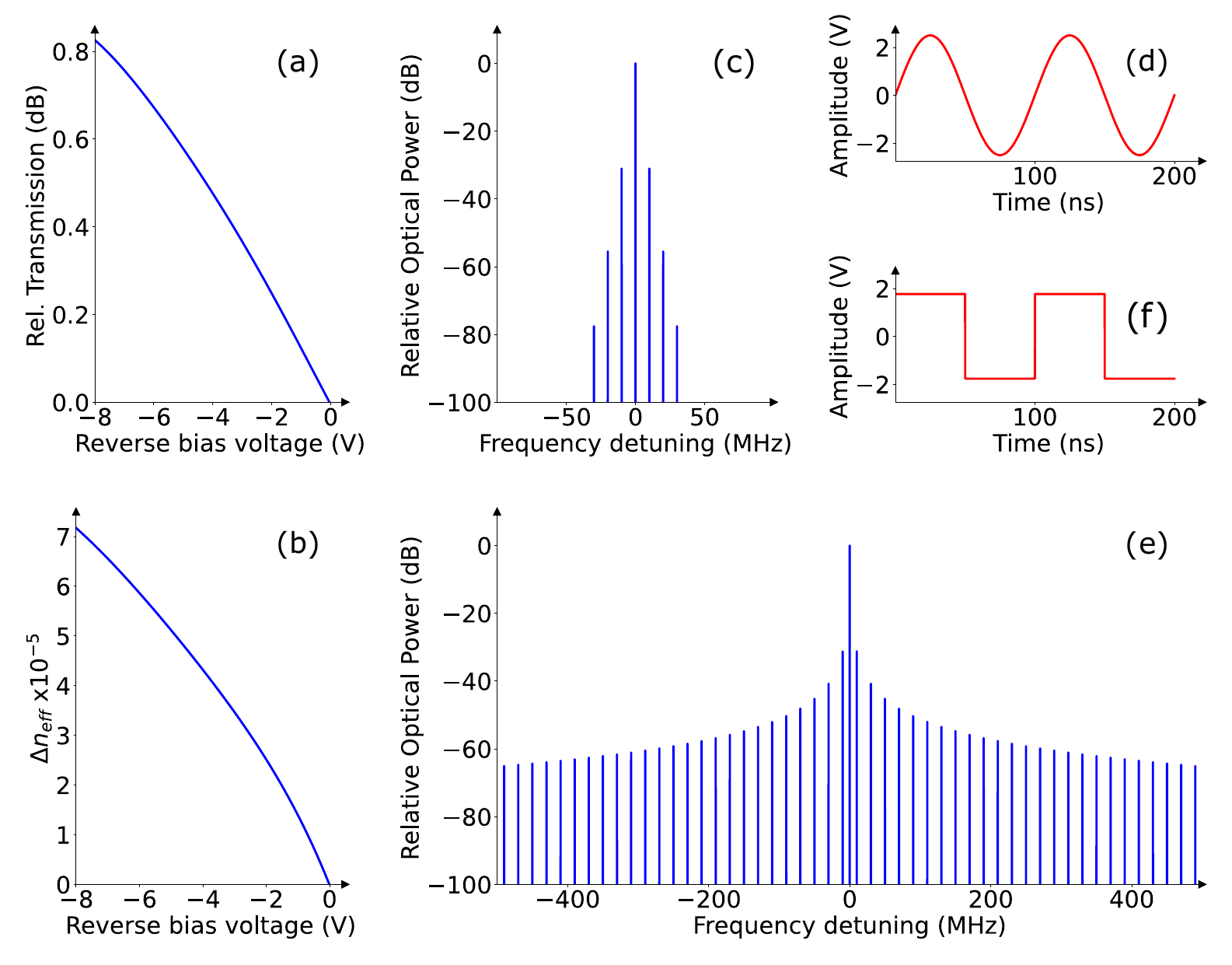}
	\caption{\textbf{Simulation of EOFC generation using the 2.2 mm long Schottky modulator. (a)} Relative optical transmission of the Schottky-based modulator on the Ge-rich graded SiGe platform, extrapolated  from the experimental measurement of a 1.7 mm long modulator at 7.8 µm wavelength reported in \cite{nguyenGHzElectroopticalSilicongermanium2022} and \textbf{(b)} simulated effective index variation at 8.5 µm wavelength, both as a function of the reverse bias voltage. These modulator characteristics are used to simulate by the mean of Equation~\ref{eq:outputLight} the output EOFC \textbf{(c), (e)} respectively  using a \textbf{(d)} 10 MHz, 5 V\textsubscript{pp}/18 dBm sinusoidal modulating signal and a \textbf{(f)} 10 MHz, 3.54 V\textsubscript{pp}/18 dBm square modulating signal.}
	\label{fig:FirstSim}
\end{figure}

\begin{figure}[ht]
	\centering
	\includegraphics[width = .9\linewidth]{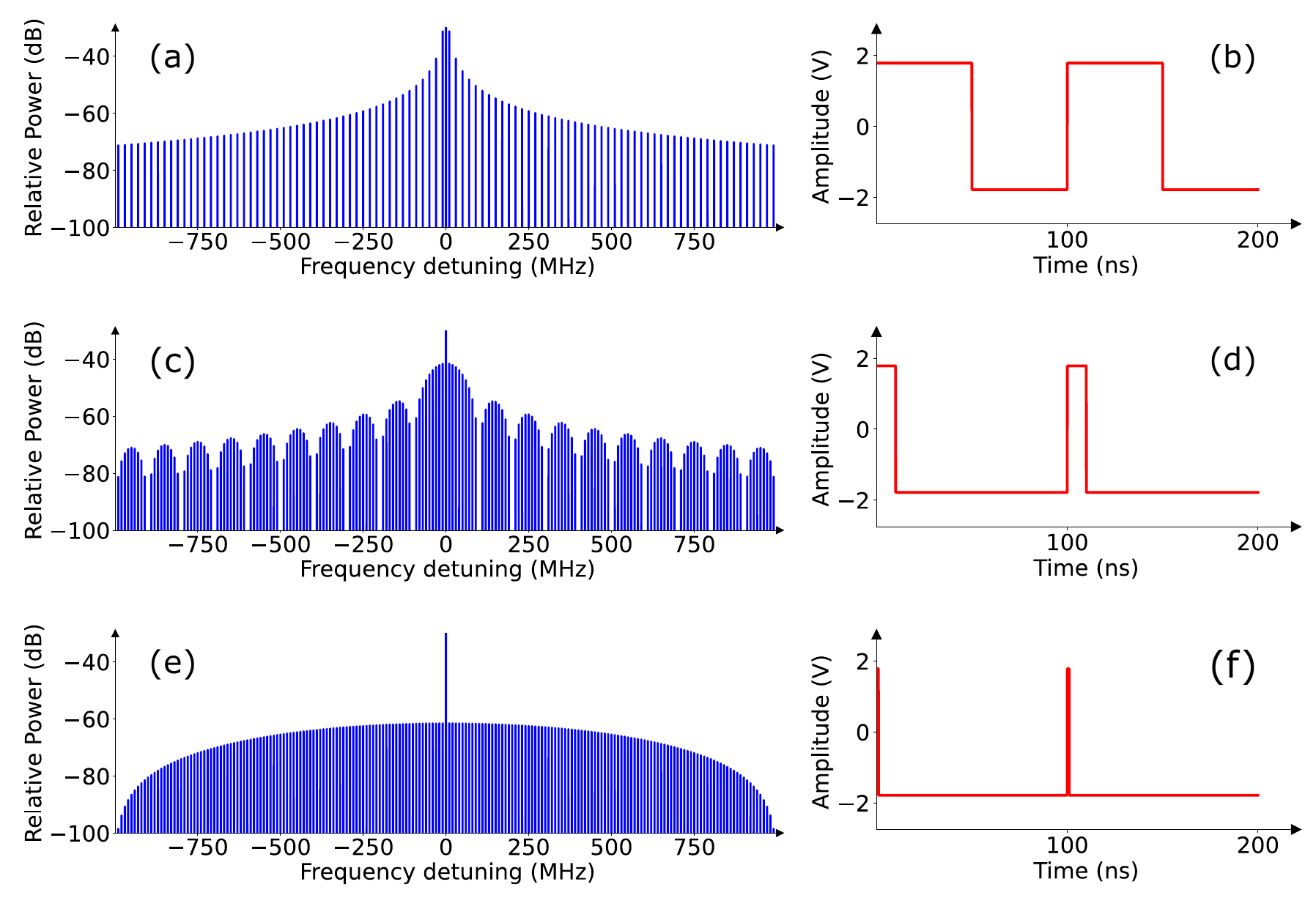}
	\caption{\textbf{Influence of the varying duty-cycle of a square signal on the output spectrum.} \textbf{(a,c,e)} Simulated output EOFC around 8 µm wavelength using 18 dBm square signals at 10 MHz respectively with \textbf{(b)} 50\% duty-cycle, \textbf{(d)} 10\% duty-cycle and \textbf{(f)} 1\% duty-cycle. The central CW line has been cropped for compactness purposes and is always at 0 dB.}
	\label{fig:compDC}
\end{figure}

The modulator used, whose architecture is reported in \cite{nguyenGHzElectroopticalSilicongermanium2022}, consists in a Schottky contact created on top of a lightly-doped Ge-rich graded SiGe structure, grown on a highly n-doped silicon substrate with a doping concentration of 2$\times$10\textsuperscript{19} cm\textsuperscript{-3}. The graded SiGe layer is 6 µm thick, with a germanium concentration that increases linearly from pure silicon at the substrate level to pure germanium. A schematic cross-sectional view of the modulator used is shown in Figure~\ref{fig:EOcombPrinciple}(a). The modulation is obtained by modulating the size of the space charge region (SCR) below the 2.2 mm long Schottky contact when applying a reverse bias, thus modifying the overlap between the represented mode and the SCR. This results in a combination of two effects, the free-carrier plasma dispersion (FCPD) which modifies the effective index of the propagating mode and the free-carrier absorption (FCA) which modifies the attenuation and consequently the output amplitude. Equation~\ref{eq:outputLight} gives a temporal expression of the output signal $A_{out}$ under these two effects.
\begin{equation}
	A_{out}(t) = A_{in}e^{i \Delta \varphi(t)}e^{-\frac{\alpha(t)}{2}L}
	\label{eq:outputLight}
\end{equation}
$A_{in}$ is the input continuous wave (CW) laser signal, $\Delta \varphi$ is the voltage-dependent phase variation due to the FCPD effect, $\alpha$ is the voltage-dependent loss due to FCA and $L$ is the modulator length. The priciple of this EOFC generation is schematised in Figure~\ref{fig:EOcombPrinciple}(b). An RF waveform is applied between the 3 top contacts designed for RF operation, shown in Figure~\ref{fig:EOcombPrinciple}(b), and the bottom contact. A CW laser beam at 8 µm is fed at the input waveguide and a frequency-comb (FC), being a set of chirped optical pulses in the temporal domain, is obtained at the output waveguide. To be able to simulate FC generation from this device, it is necessary to estimate the phase and loss variations as a function of the reverse bias applied around 8 µm wavelength. Previous measurements \cite{nguyenGHzElectroopticalSilicongermanium2022} have been used to estimate the relative transmission variation of the 2.2 mm long modulator as a function of the applied voltage, as reported in Figure~\ref{fig:FirstSim}(a). Numerical simulations of the bias-dependent effective index variation have also been performed and the results are represented in Figure~\ref{fig:FirstSim}(b). These quantities respectively allow us to evaluate $\alpha(t)$ and $\Delta \varphi(t)$, hence the output power spectral density (PSD) is calculated taking the modulus squared of the Fourier transform of Equation~\ref{eq:outputLight}, assuming that the mismatch of electrical and optical velocities and RF wave attenuation are negligible for 2.2 mm long electrodes~\cite{denielSiliconPhotonicsPhase2021}. It is important to note that as the effective index variation is limited, the amplitude modulation is the dominating effect.
The results of the numerical simulation of the output EOFC from a 2.2 mm long Schottky modulator, assuming a purely monochromatic input at 8 µm wavelength and a 5 V\textsubscript{pp} sinusoidal signal at 10 MHz frequency are shown in Figure~\ref{fig:FirstSim}(c). 
As the amplitude modulation is mainly in the linear regime due to the low absorption variation, the spectral width of the output EOFC is limited, by the spectral width of the RF modulating signal used. This makes the use of harmonically rich modulating signals a possible candidate to increase the number of generated lines from a single modulator, since as long as the all the harmonics of the signal are contained in the reported 1 GHz modulator bandwidth \cite{nguyenGHzElectroopticalSilicongermanium2022}, the number of generated lines will increase with the number of harmonics of the modulating signal. As an insight, the Figure~\ref{fig:FirstSim}(e) shows the simulated output EOFC from the same modulator, with the same input optical and electrical powers as in Figure~\ref{fig:FirstSim}(c), but using a square modulating signal. It could be observed that this drastically increases the number of comb lines. This used square signal has a 50 \% duty-cycle, leading to a \textit{sinc-like} spectral envelope with 20 MHz wide side lobes and a linespacing being the repetition frequency of the square signal, ending up in suppressing half of the generated lines. This problem could be overcome by adjusting the duty-cycle of the modulating signal, as shown in Figures~\ref{fig:compDC}(a), \ref{fig:compDC}(c) and \ref{fig:compDC}(e), which are the simulated output EOFC when using the modulating signal respectively shown in Figures~\ref{fig:compDC}(b), \ref{fig:compDC}(d) and \ref{fig:compDC}(f).
Several observations arise from this. First, this duty-cycle reduction inevitably comes with a trade-off. An extremely small duty-cycle produces an extremely large and flat central spectral lobe, but the output optical signal will be mainly CW. The power of the spectral component at the optical carrier frequency is going to be orders of magnitude above the one of the generated lines. Secondly, the $\frac{1}{f}$ decay of the spectral envelope of a square signal implies that there are many remaining harmonics above the modulator bandwidth, meaning that a non-negligible portion of the electrical power is going to be filtered out by the modulator.
\begin{figure}[ht!]
	\centering
	\includegraphics[width=.9\linewidth]{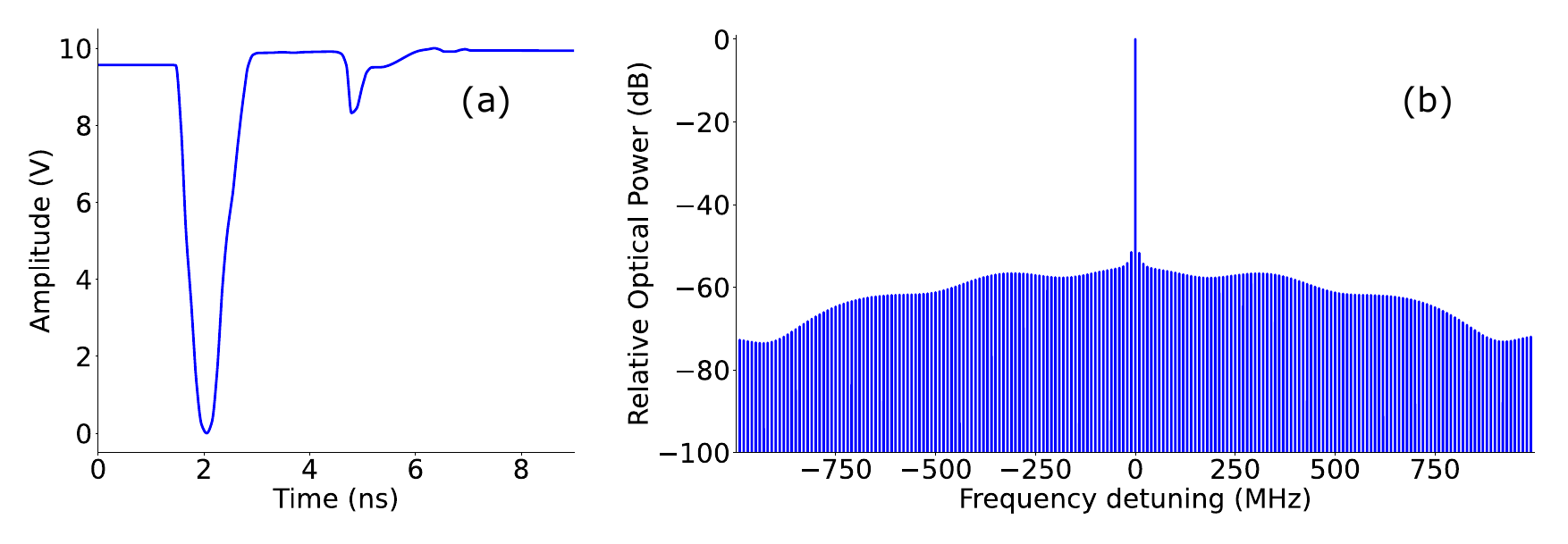}
	\caption{\textbf{EOFC generation using the RF pulse generator. (a)} Output pulse of the pulse generator (UltraComb-8G, Ultraview). Its FWHM is 739 ps with a repetition frequency of 10 MHz, with a 10 V\textsubscript{pp} amplitude. From simulation, we can expect it to generate the EOFC shown in \textbf{(b)} when it is used to drive the modulator.}
	\label{fig:pulse}
\end{figure}
Moreover, the generation of such square voltage pulses requires an arbitrary waveform generator (AWG), which is a bulky and either expensive or relatively slow instrument. In order to push this idea further, we propose the use of a pulse generator as driving signal for the modulator. As a matter of fact, an RF pulse generator is an RF frequency-comb generator, whose envelope and linespacing can be tuned by respectively tuning the shape of the pulse and the number of pulse per second. The shape of these pulses has also the advantage of having a limited spectral spanning and allow to maximize the power transfer to the modulator. The pulse generator (UltraComb-8G, Ultraview) used in the following has been set to emit \mbox{10 V\textsubscript{pp}}, 739 ps full-width at half-maximum (FWHM) pulses, with a 10 MHz repetition frequency. The measured output pulse is shown in Figure~\ref{fig:pulse}(a), and the simulated output optical spectrum using these pulses to drive the modulator is shown in Figure~\ref{fig:pulse}(b).

\section{Experiments}
\subsection{Experimental setup}
\begin{figure}[ht]
	\centering
	\includegraphics[width=.7\linewidth]{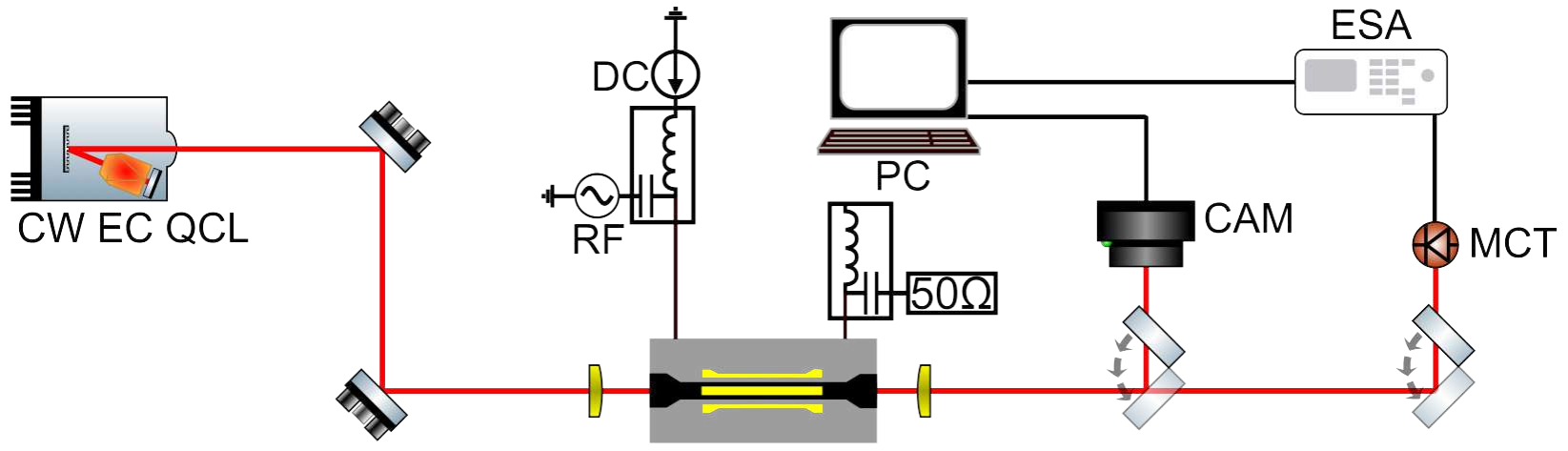}
	\caption{\textbf{Simplified version of the experimental setup used}. The output of a CW external cavity QCL (\textbf{CW EC QCL}) is coupled to the chip using aspheric ZnSe lenses, and collected similarly before being sent to either a mode profiler (\textbf{CAM)} or a fast pre-amplified MCT detector (\textbf{MCT}), whose output is analyzed using an electrical spectrum analyzer (\textbf{ESA}). The modulator is polarised by a SMU (\textbf{DC}) and driven either by an AWG or an ultrafast RF comb generator (\textbf{RF}).}
	\label{fig:ExpSetup}
\end{figure}

In order to characterize the output EOFC, laser emission coming from a CW QCL emitting around 8 µm wavelength (MIRcat, Daylight Solutions) is coupled in the modulator using a pair of aspheric ZnSe lenses, and sent either to a mode profiler (WinCamD IR-BB, DataRay) to ensure a correct coupling, or to a pre-amplified fast MCT detector (UHSM-I-10.6, Vigo System), whose electrical output is analysed using an electrical spectrum analyzer (MS2830A, Anritsu). The specified MCT detector electrical bandwidth is 700 MHz. In order to apply the modulating signal to the 2.2 mm long modulator, an RF signal either coming from an arbitrary waveform generator (AWG 3390, Keithley) or an RF pulse generator (Ultracomb-8G, Utraview) is applied to the Schottky modulator while using a bias-tee and a source measure unit (SMU 2401, Keithley) to ensure that the modulator is always operating in the reverse regime. Another RF probe with a 50$\Omega$ load is used on the other side of the travelling wave electrode (TWE) of the modulator. A schematic view of the experimental setup used is given in Figure~\ref{fig:ExpSetup}.
It is of major importance to note that using this measurement scheme, the generated EOFC will produce a signal at the output of the MCT detector consisting in the beating between the different generated lines. Indeed, the presence of a beatnote at a frequency $\nu_{RF}$ only proves that there are at least 2 lines spaced by $\nu_{RF}$ in the frequency domain. This beatnote at $\nu_{RF}$ is thus coming from the sum of the beatnotes produced by all the frequencies separated by $\nu_{RF}$. However, the detected signal from the beatnote between two lines is proportional to the product of the corresponding Fourier coefficients in the series development of the expression of the EOFC. This implies that, the carrier line being expected to be orders of magnitude above the generated comb lines, the main origin of a beatnote at $\nu_{RF}$ is the beating between the optical carrier line and the two lines distant from this carrier by $\nu_{RF}$. In such a case, it is reasonable to consider that this beatnote measurement is a good approximation of the sum of the two side of the generated symmetric spectrum.

\subsection{Results}
\begin{figure}[ht]
	\centering
	\includegraphics[width=.69\linewidth]{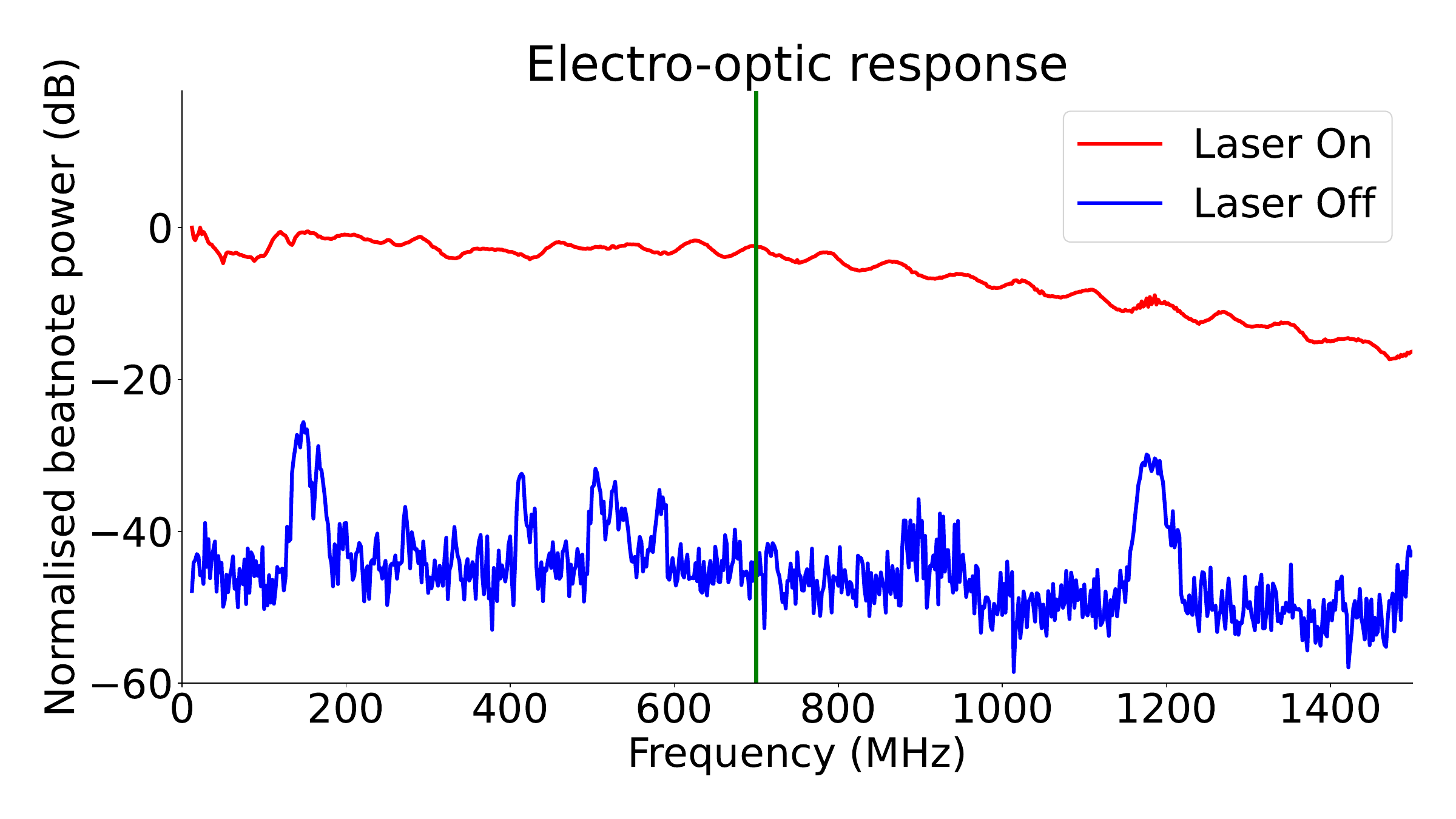}
	\caption{\textbf{Normalised electro-optic response of the modulator used for the EOFC experiments}. The beatnote power is shown in \textbf{red} when the laser is on and in \textbf{blue} when it is off. To emphasise the origin of the drop in the beatnote power, the cut-off of the pre-amplifier of the MCT photodetector is plotted in \textbf{green}.}
	\label{fig:EOresponse}
\end{figure}

\begin{figure}[ht]
	\centering
	\includegraphics[width=.8\linewidth]{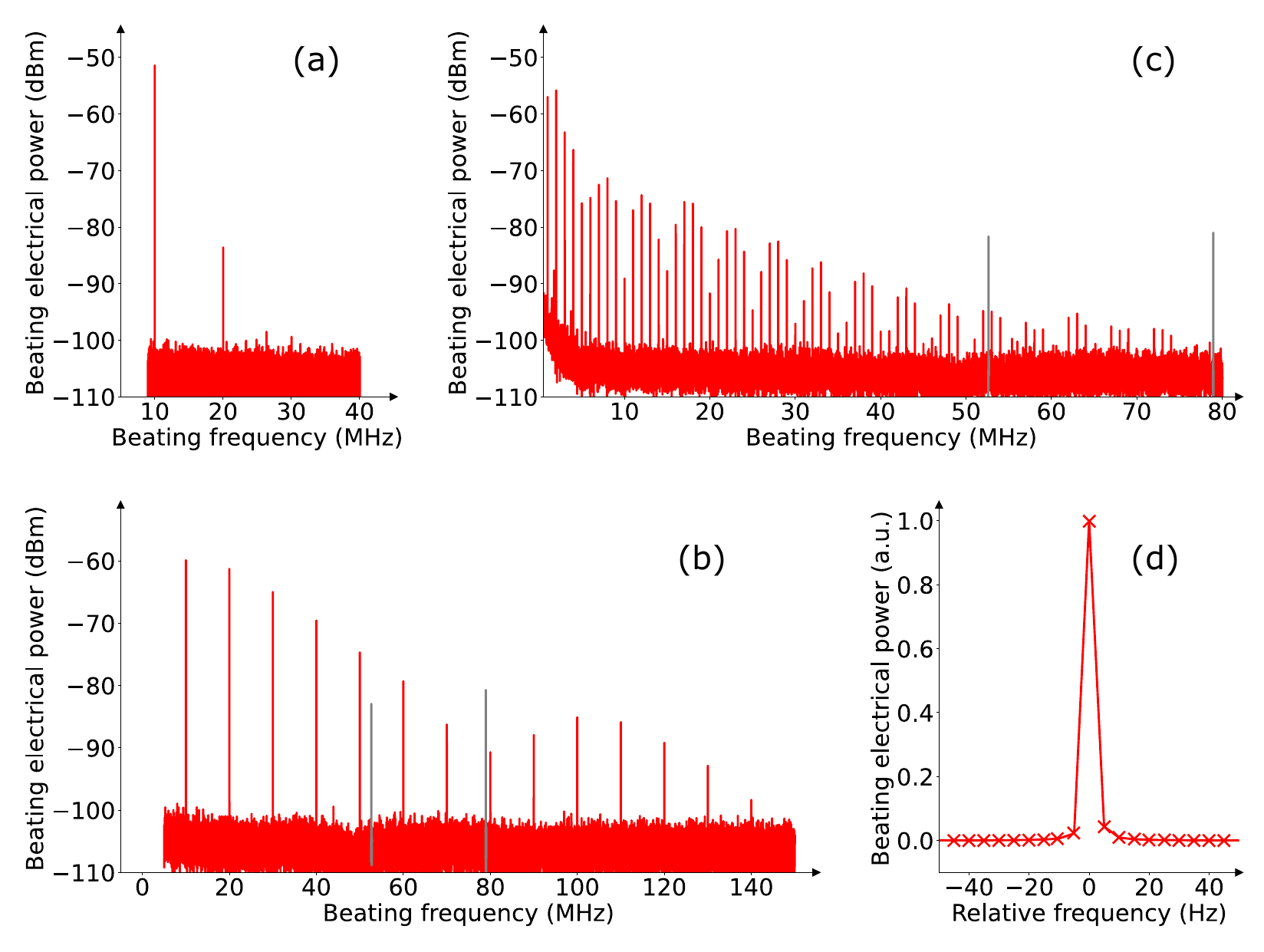}
	\caption{\textbf{Experimental demonstration of EOFC generation at 8 µm wavelength. (a)} Beatnote signal coming from the generated EOFC using a CW laser at 8 µm wavelength and a 5 V\textsubscript{pp}, 10 MHz sinusoidal signal to modulate. \textbf{(b)} Beatnote signal measured under the same conditions but using a 20\% duty-cycle, 4 V\textsubscript{pp} square signal. Some parasitic lines, emphasised in grey and coming from the setup could be observed around 53 and 79 MHz. \textbf{(c)} Beatnote signal under the same conditions as in \textbf{(b)}, but using a 1 MHz repetition rate. \textbf{(d)} To emphasise the sharpness of the comb lines, a zoom on the first line of the beatnote signal shown in (a) is performed. A FWHM of the beatnote line of 5 Hz is measured, limited by the Doppler resolution, the acquisition time being 0.2 seconds.}
	\label{fig:ExpSinvsSq}
\end{figure}

\begin{figure}[ht]
	\centering
	\includegraphics[width = .8\linewidth]{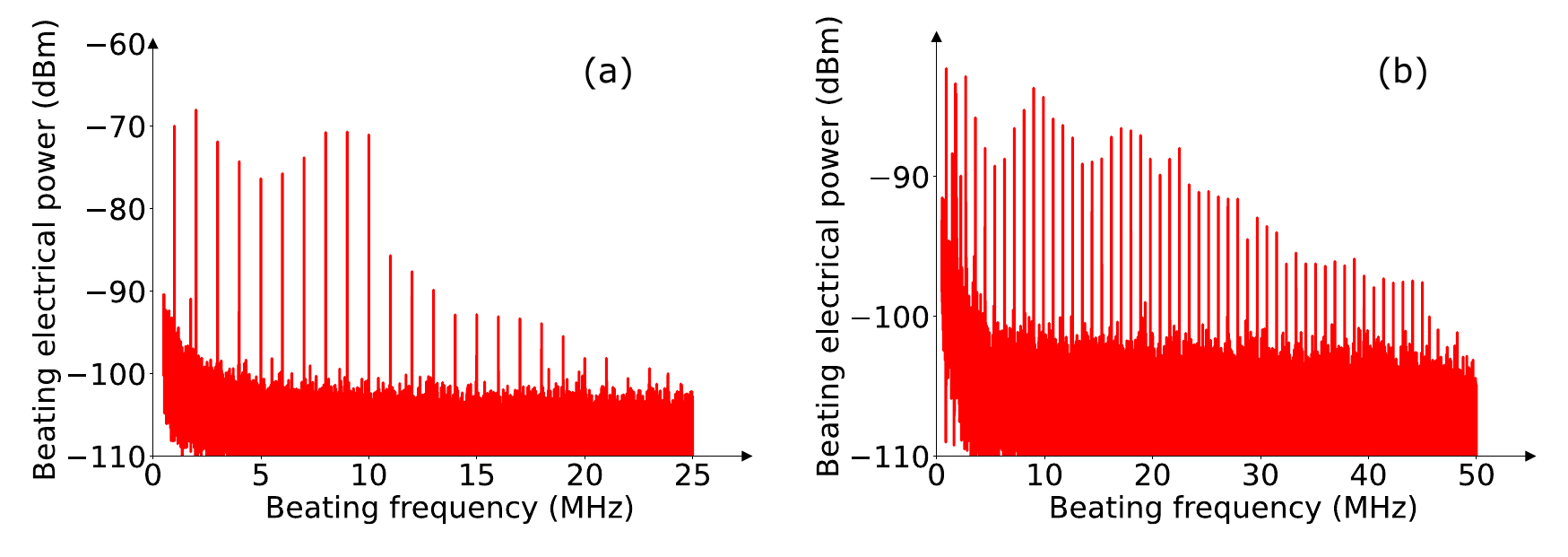}
	\caption{\textbf{Using \textit{sinc}-shaped signals to generated EOFC.} Resulting beatnote of the generated EOFC using the signal whose expression is given in Equation~\ref{eq:diric} for a 1 MHz repetition rate and an amplitude of 5 V\textsubscript{pp}, with \textbf{(a)} $n=21$ and \textbf{(b)} $n=101$.}
	\label{fig:sinc_shaped_exp}
\end{figure}

\begin{figure}[!ht]
	\centering
	\includegraphics[width=.8\linewidth]{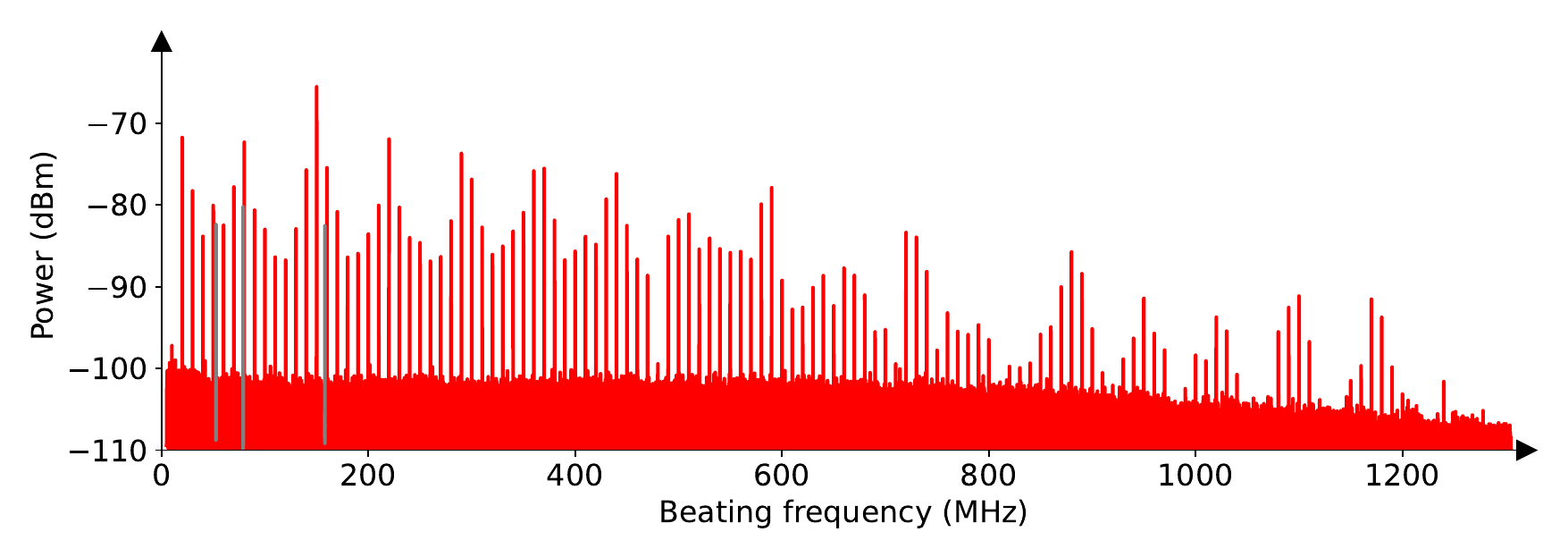}
	\caption{Measured beatnote using the driving signal shown in Figure~\ref{fig:pulse}(a). The signal to noise ratio drops after 700 MHz both because of the fast MCT cut-off emphasised in Figure~\ref{fig:EOresponse} and the theoretical amplitude of the lines dropping after 750 MHz of detuning with respect to the optical carrier line, as shown in Figure~\ref{fig:pulse}(b). The parasitic lines coming from the setup are emphasised in grey.}
	\label{fig:PulseExp}
\end{figure}

First, we verify that the modulator used possesses a similar speed to the one reported in \cite{nguyenGHzElectroopticalSilicongermanium2022}. To do so, we apply an RF signal using an RF synthesiser (MG3694C, Anritsu), whose power is fixed and whose frequency is swept from 10 MHz to 1.5 GHz, to the modulator and measure the beatnote power at the modulation frequency on the electrical spectrum analyzer (ESA). The resulting EO response is shown in Figure~\ref{fig:EOresponse}.
One can observe that the EO response drops after 700 MHz, because of the cut-off of the pre-amplifier of the fast MCT photodetector, which is going to reduce the signal to noise ratio after this cut-off frequency. However, no sharp drop is observed, indicating that the modulator can work up to 1.5 GHz at least. This will allow us to perform some experiments beyond the reported 1 GHz EO bandwidth of this modulator architecture.\\
We then demonstrate the ability of this modulator to generate EOFC. To do so, we apply using the AWG a 10 MHz, 5 V\textsubscript{pp} sinusoidal signal on the modulator and measure the beatnote spectrum output by the MCT detector on the ESA, while setting up the CW EC QCL at 8 µm wavelength. The resulting spectrum is shown in Figure~\ref{fig:ExpSinvsSq}(a). One can observe that only two beatnotes are observed at 10 and 20 MHz, indicating that applying a sinusoidal modulating signal may lead to only 2 measurable lines on each side of the optical carrier. However, when applying using the same AWG a 4 V\textsubscript{pp}, 20\% duty-cycle square signal on the modulator at the same frequency, we obtain the beatnote signal reported in Figure~\ref{fig:ExpSinvsSq}(b) which, apart from some parasitic lines emphasised in grey coming from the setup, possesses many more lines, indicating that a much wider EOFC, limited by the AWG bandwidth, is being generated. The linespacing of such a FC could be freely tuned by adjusting the repetition rate of the modulating signal. As an example, Figure~\ref{fig:ExpSinvsSq}(c) shows the measured beatnote signal under the same conditions as in Figure~\ref{fig:ExpSinvsSq}(b), except the repetition rate fixed to 1 MHz. It is also worth noting that the \textit{sinc-like} shape of the envelope of the spectrum is obtained as expected, even though only the interline beatnote is measured. This shows as well that we can freely adjust both the shape and the linespacing of the generated EOFC by changing the modulating waveform and the repetition rate. In order to show the precision spectroscopy abilities of such an EOFC, a precision measurement of the linewidth of the beatnote lines is performed using a 12-bit digitizing acquisition card (ATS9371, AlazarTech) and performing a Fourier transform of the 0.2 seconds long acquisition at 1 GS/s sampling rate, in the same condition as in Figure~\ref{fig:ExpSinvsSq}(a). The result of this measurement is shown in Figure~\ref{fig:ExpSinvsSq}(d). We can observe that the measured linewidth is Doppler-limited at least for any acquisition time below 200 ms, since the measured FWHM of the line is 5 Hz.  
The ability to create any desired spectral shape is unlocked by the possibility to synthesise any waveform electronically. Since flatness is a relevant figure of merit in FC generation, it is natural to consider the use of \textit{sinc}-shaped modulating signal in order to achieve a flatter output spectrum. We thus applied a \mbox{1 MHz}, 5 V\textsubscript{pp} periodic \textit{sinc} signal, also called a Dirichlet function, whose expression is given in Equation~\ref{eq:diric} to the modulator,
\begin{equation}
	S(\varphi,n) \propto 1-\frac{sin\left(\varphi.\frac{n}{2}\right)}{n.sin\left(\frac{\varphi}{2}\right)},
	\label{eq:diric}
\end{equation}
where $\varphi$ is the instantaneous phase and $n$ the order of the signal. The higher the value of $n$, the higher the number of oscillations that are going to occur within a period resulting in a sharper peak and a more extended spectrum. This signal has been used to modulate for $n=21$ and $n=101$, and the corresponding spectra of the beatnotes are respectively given in Figures~\ref{fig:sinc_shaped_exp}(a) and \ref{fig:sinc_shaped_exp}(b).
It is worth noting that even though the $\frac{n}{2}-1$ lines are supposed to be flat for an order $n$ of the modulating signal given in Equation~\ref{eq:diric}, we observe that the output spectrum has its envelope affected by the EO transmission of the modulator reported in Figure~\ref{fig:EOresponse}, as well as by the AWG bandwidth limitation at 10 MHz for arbitrary waveforms. This leads to ripples and amplitude drops in the output EOFC, but the output of the system could still be predicted knowing these responses.\\
To go a step further towards fully integrated EOFC sources, we also demonstrate the possibility to generate an EOFC spanning over a total of more than 2 GHz with a 10 MHz linespacing using the electrical pulse generator. Its output pulses are similar to the ones that could be generated using an integrated synthesiser followed by a step recovery diodes based impulse train generator module (\textit{e.g.} 33002 A, HP). Both are compact and CMOS compatible devices that could be integrated and used to drive a modulator. The pulse generator whose output is shown in Figure~\ref{fig:pulse}(a) is thus used to drive the modulator, and the corresponding beatnote signal measured is given in Figure~\ref{fig:PulseExp}.
This demonstrates the ability to generate EOFC spanning over 2.4 GHz / 500 pm around the optical carrier line at 8 µm wavelength, using palm-sized driving electronic systems and a single Schottky modulator thanks to the silicon-compatible Ge-rich graded SiGe platform. Knowing that the electrical pulse generator is adjustable in repetition frequency and pulse width, this paves the way towards fully integrated EOFC sources that are tunable in repetition rate and spectral envelope, by simply turning a knob.

\section{Conclusion}
In conclusion, EOFC generation has been demonstrated for the first time in the LWIR spectral range, using a Schottky modulator in the graded SiGe platform. A very small linewidth of 5 Hz has been demonstrated together with the ability to tune the repetition rate by simply tuning the repetition rate of the applied RF signal. It was also shown that using an integration-compatible RF pulse source, an EOFC spanning over 2.4 GHz / 500 pm could be generated around 8 µm wavelength. These results unlock the use of compact photonics circuits for on-chip fine resolution spectroscopy. Furthermore, it is worth noting that the used modulator has shown operation in a wide spectral range, from 5 to 9 µm wavelength \cite{nguyenGHzElectroopticalSilicongermanium2022}, so the central wavelength of the generated comb is also widely tunable. Further developments on the bandwidth and modulation depth of the electro-optical modulators would lead to an increase of the energy conversion and a larger number of generated lines.



\medskip
\textbf{Supporting Information} \par 
Data underlying the results presented in this paper may be obtained from the authors upon reasonable request.

\medskip
\textbf{Acknowledgements} \par 
This work was supported by ANR Light-Up Project (No.
ANR-19-CE24-0002-01). The fabrication of the device was partially performed within the Plateforme de Micro-Nanotechnologie/C2N, partially funded by the "Conseil Départemental de l'Essonne". It was partly supported by the french RENATECH network.

\medskip
\textbf{Disclosure} \par
The authors declare no conflict of interest.
\medskip
%
\bibliographystyle{unsrt}
\bibliography{myBib}

\end{document}